\documentclass[twocolumn,aps,prd,showpacs,reprint,longbibliography,nofootinbib,ngerman,UKenglish]{revtex4-1}
 \usepackage{natbib}
 
 \usepackage[utf8x]{inputenc}
 \usepackage{braket,amsmath,mathrsfs,units}
 \usepackage{amsthm,amsfonts}
 \usepackage[locale = DE]{siunitx}
 \usepackage{dsfont}
 \usepackage{amssymb,fixmath}
 \usepackage{caption,subcaption}
 
 \usepackage{hyperref}
 \usepackage{enumerate}
 
 \usepackage{appendix}
 \usepackage[main=newzealand, greek]{babel}
 \usepackage[LGR,T1]{fontenc}
 \usepackage{csquotes}
 \usepackage{tcolorbox}
 \usepackage{calc}
 \usepackage{nameref}
 \makeatletter
 \newcommand*{\currentname}{\@currentlabelname}
 \makeatother
 
 \allowdisplaybreaks 
 
 \newcommand{\abs}[1]{\left\lvert #1 \right\rvert}
 \newcommand{\dif}{\ensuremath{\mathrm{d}}}
 \newcommand{\kl}[1]{\left( #1 \right)}
 
 \newcommand{\kle}[1]{\left[ #1 \right]}
 \newcommand{\ed}[1]{\frac{1}{#1}}
 \newcommand{\defi}{\mathrel{\mathop:}=}
 \newcommand{\ifed}{=\mathrel{\mathop:}}
 \newcommand{\eps}{\varepsilon}

 \newcommand{\mi}{\ensuremath{{\left[\mu^{-1}\right]}}}
 \newcommand{\zt}{\kle{\zeta^\dagger}}
 \newcommand{\gi}{\kle{g_{\text{eff}}^{-1}}}
 \newcommand{\g}{g_{\text{eff}}}
 \newcommand{\pdet}{\ensuremath{\mathrm{pdet}}}

\begin{document}
 	\title{Effective metrics and a fully covariant description of constitutive tensors in electrodynamics}
 	\author{Sebastian Schuster}
 	\email{sebastian.schuster@sms.vuw.ac.nz}
 	\author{Matt Visser}
 	\email{matt.visser@sms.vuw.ac.nz}
 	\affiliation{School of Mathematics and Statistics,\\
 		Victoria University of Wellington; \\
 		PO Box 600, Wellington 6140, New Zealand.}
 	\date{\today}
 		
 \begin{abstract}
 		Using electromagnetism to study analogue space-times is tantamount to considering consistency conditions for when a given (meta-)material would provide an analogue space-time model or --- vice versa --- characterizing which given metric could be modelled with a (meta-)material. 
While the consistency conditions themselves are by now well known and studied, the form the metric takes once they are satisfied is not. This question is mostly easily answered by keeping the formalisms of the two research fields here in contact as close to each other as possible. While fully covariant formulations of the electrodynamics of media have been around for a long while, they are usually abandoned for (3+1)- or 6-dimensional formalisms. Here we shall use the fully unified and fully covariant approach. This enables us even to generalize the consistency conditions for the existence of an effective metric to arbitrary background metrics beyond flat space-time electrodynamics. We also show how the familiar matrices for permittivity $\epsilon$, permeability $\mu^{-1}$, and magneto-electric effects $\zeta$ can be seen as the three independent pieces of the Bel decomposition for the constitutive tensor $Z^{abcd}$, i.e., the components of an orthogonal decomposition with respect to a given observer with four-velocity $V^a$. 
Finally, we shall use the Moore--Penrose pseudo-inverse and the closely related pseudo-determinant to then gain the desired reconstruction of the effective metric in terms of the permittivity tensor $\epsilon^{ab}$, the permeability tensor $\mi^{ab}$, and the magneto-electric tensor $\zeta^{ab}$, as an explicit function $\g(\epsilon,\mu^{-1},\zeta)$.	

\vspace*{.3cm}
\noindent\href{https://arxiv.org/abs/1706.06280}{arXiv:1706.06280},\qquad Phys.~Rev.~D \textbf{96}:124019(2017), \qquad DOI: \href{https://doi.org/10.1103/PhysRevD.96.124019}{10.1103/PhysRevD.96.124019}
\end{abstract}

 	\maketitle
 	
 	\section{Introduction}
 	When studying analogue space-times \cite{lrrAnalogue,TrafoOptics2}, one has a choice of many different approaches. Basically whenever a physical model, or approximation thereof, provides a wave equation for some scalar physical quantity $f$, one can look for a metric $\g$ such that this wave equation would be the corresponding Laplace--Beltrami equation
 	\begin{equation}
 		\nabla_a\nabla^a f = f_{;a}{}^{;a} = \ed{\sqrt{\abs{\det \g}}} \partial_a \kl{\sqrt{\abs{\det \g}}\;\partial^a f} = 0,
 	\end{equation}
 	though maybe an inhomogeneous one. More generally, (as in the present case of electrodynamics), a general wave equation is a Lorentz-invariant\footnote{Though not necessarily with respect to the speed of light in vacuum!}, hyperbolic partial differential equation (PDE) of second order. Likewise, instead of the Laplace--Beltrami equation one wants to express this given PDE as another PDE of the same type, but now depending on an effective, Lorentzian metric $\g$ and its (Lorentzian) geometry. The abundance of (tensorial) wave equations in physics therefore raises the question of when exactly this can be done. In this paper, we shall take a look at the special case of (four-dimensional) macroscopic electrodynamics, i.e., electrodynamics in a medium. In this particular context the question also quickly becomes one of finding an appropriate formalism: If we want to study a given wave equation through an analogue space-time a fully covariant approach will prove to be the most natural approach. But while microscopic electrodynamics (i.e., electrodynamics in vacuum albeit with sources) in flat space easily provides the well-known, fully covariant Maxwell equations\footnote{See, for example, references \cite{GourSR} or \cite{MTW}.}
 		\begin{equation}
	 		\partial_{[a} F_{bc]} = 0, \qquad F^{ab}{}_{;b} = \epsilon_0^{-1} j^a,
	 	\end{equation}
 	this fully covariant approach is a bit more intricate in the context of electrodynamics in media. While results have been known for a long time, see for example \cite{BalZim,BalNi,AreaMetrClass,AniBiGuide,Post,Marx1,Schmutzer,HehlLaemmerzahl,FavaroBergamin} and \cite{ODell}, they have rarely been used to full extent. The general idea is to exchange the metric dual of the field strength tensor
 	\begin{equation}
	 	F^{ab} =  g^{ac} g^{bd} F_{cd} = \ed{2} \kl{ g^{ac} g^{bd} -  g^{ad} g^{bc}}F_{cd}\label{eq:ZBG}
 	\end{equation}
 	with the excitation tensor
 	\begin{equation}
 		G^{ab} \defi Z^{abcd} F_{cd}.\label{eq:excitationtensor}
 	\end{equation}
 	Here $Z^{abcd}$ is the so-called constitutive tensor (or general susceptibility tensor \cite{ODell}). What is usually done is to use the properties of the constitutive tensor (elaborated below) and switch from four space-time indices $a, b, c, \dots$ ranging from 0~to~3 to two \enquote{field indices} $A, B, C, \dots$ ranging from 1~to~6. This enables one to collect the index pair $ab$ into a new compound index $A$ and the index pair $cd$ into a new compound index $B$. Schematically:
 	\begin{equation}
 		\kl{Z^{abcd}}_{a,b,c,d\in\{0,\dots, 3\}} \to \kl{Z^{AB}}_{A,B\in\{1,\dots, 6\}}.\label{eq:6x6}
 	\end{equation}
 	The issue here is that one loses the full covariance and instead implicitly uses an observer-dependent 3+1 decomposition. In the context of pre-metric electrodynamics (see, for example, \cite{HehlObukhov} and references therein) this is not a bug, but a feature. Our current approach is orthogonal to the pre-metric one: Not only do we want to keep the physical background metric $g$, we will also look for an effective metric $\g$. As both metrics will be four-dimensional and general, we want to stick with space-time indices.
 	
 	As a result, the strategy in this paper is two-fold: First, we want to showcase this fully covariant formalism for electrodynamics of media using only space-time indices as it is done, for example, in \cite{PerlickRay}, and \cite{BalZim}. Second, we also want to find the consistency conditions in terms of the constitutive tensor that have to be fulfilled in order for it to describe a material providing a model for analogue space-times. The requirement for this to work is that the constitutive tensor $Z$ can be written in terms of an effective metric $\g$, analogously to equation~\eqref{eq:ZBG}, as
 	\begin{equation}
 		Z^{abcd} = \ed{2}\frac{\sqrt{\det \g}}{\sqrt{\det  g}} \kl{\gi^{ac}\gi^{bd} - \gi^{ad}\gi^{bc}}.\label{eq:Zeff}
 	\end{equation}
 	As this second point in turn is important when engineering materials for this purpose, we shall give these consistency conditions in terms of the familiar matrices $\epsilon, \mu^{-1}$, and $\zeta$ (or their four-dimensional generalisations).
 	
While the derivation of the consistency conditions has been done before (in numerous and various contexts and formalisms), see for instance references~\cite{BalZim,BalNi,AreaMetrClass,HehlLaemmerzahl,FavaroBergamin}, it still remains to explicitly write down the resulting effective metric once the consistency conditions are satisfied. (In the context of pre-metric electrodynamics this is quite naturally done as soon as  the spacetime metric is recovered \cite{HehlLaemmerzahl,HehlObukhov,RubilarPremetric}.)

\enlargethispage{20pt}

However, as we shall work assuming a non-trivial background metric $g_{ab}$ for the material, the approach herein differs greatly. We shall soon see that, whenever an electromagnetic medium can be characterized by an effective metric $[\g]_{ab}$, there always exists an observer with four-velocity $V^a$ in whose rest frame the well-known consistency conditions 
 	\begin{equation}
 	 	\epsilon^{ab} = \mu^{ab}, \qquad \zeta^{ab} = 0
 	\end{equation}
 	hold. 
In this rest frame the effective metric $\g$ can be written in terms of Moore--Penrose pseudo-inverse and the related pseudo-determinant in the following way:
	\begin{subequations}
	 	\begin{align}
	 	 	\kl{g_\mathrm{eff}}_{ab} &= -\sqrt{\frac{-\det(g^{\bullet\bullet})}{\pdet(\epsilon^{\bullet\bullet})}}\, V_a V_b + \sqrt{\frac{\pdet(\epsilon^{\bullet\bullet})}{-\det(g^{\bullet\bullet})}} [\epsilon^{\bullet\bullet}]^\#_{ab},\\
	 	 	&= -\sqrt{\frac{-\det(g^{\bullet\bullet})}{\pdet(\mu^{\bullet\bullet})}}\, V_a V_b + \sqrt{\frac{\pdet(\mu^{\bullet\bullet})}{-\det(g^{\bullet\bullet})}} [\mu^{\bullet\bullet}]^\#_{ab}.
	 	\end{align}
	\end{subequations}
 	
 	The paper is organized as follows: First we recapitulate the properties of the constitutive tensor, also elaborating a bit on the traditional rewriting as $Z^{AB}$. In the second section we shall develop from this a 3+1 decomposition of the constitutive tensor. For a given observer moving with some arbitrary four-velocity $V^a$, this provides the link between electric field $\mathbf{E}$, magnetic field $\mathbf{B}$ and the corresponding displacement field $\mathbf{D}$ and magnetizing field $\mathbf{M}$ via the \enquote{constitutive matrices} $\epsilon, \mu^{-1}$, and $\zeta$. In this (3+1)-decomposed case we shall derive the sought-after consistency conditions. Having done so it is then possible to develop the next section: The fully covariant formulation and the corresponding version of the consistency conditions. After concluding, we provide two appendices: First, a small aside on the relation between the constitutive tensor's Bel decomposition and the $\epsilon, \mu^{-1}$, and $\zeta$ three-tensors, and second, an example application of the formalism presented here to the case of moving, isotropic media.
 	
 	\section*{Notation}
 	This paper follows the sign conventions of \cite{GourSR} and \cite{MTW}. Specifically, our metrics will have signature $(-+++)$. Symmetrisation and antisymmetrisation on indices is indicated by enclosing these indices in round or square brackets, respectively. Raising and lowering of indices shall always be done employing the physical background metric $ g$. For the sake of brevity, we shall not always place \enquote{physical} before \enquote{background metric}. If indices need to be raised or lowered by $\g$, $\g$ shall appear explicitly. $\bullet$ (for four-indices) and $\circ$ (for three-indices) are used to denote index placement, mostly used in determinants. We use the symbol $\stackrel{!}{=}$ whenever we manually set things equal or demand them to be equal.
 	
 \section{General properties of the constitutive tensor}
 \subsection{Counting degrees of freedom}
 	A first part of the analysis is to compare the degrees of freedom of the effective metric and the constitutive tensor. 

\noindent	
Note that quite generally the action in our case will be
 	\begin{align}
	 	S &= -\int \dif^4 x \frac{\sqrt{-\det  g}}{4} F_{ab} G^{ab},\\
	 	&= -\int \dif^4 x \frac{\sqrt{-\det  g}}{4} F_{ab} Z^{abcd} F_{cd}, \label{eq:action}
	\end{align}
	plus possible source terms. Assuming the existence of an effective metric then enforces 
	\begin{align}
	 	S \stackrel{!}{=}& -\ed{8} \int \dif^4x \sqrt{-\det \g}\nonumber\\ &\times \kl{\gi^{ac}\gi^{bd} - \gi^{ad}\gi^{bc}}F_{ab}F_{cd}.\label{eq:actioneff}
	\end{align}
 	From this it follows that the action is invariant under conformal transformations of $\g$. Instead of the regular degrees of freedom of a symmetric $4\times4$ matrix, $\g$ therefore has only $\nicefrac{4(4+1)}{2}-1 = 9$ degrees of freedom.
 	
 	For the degrees of freedom of $Z$, again take a look at equation~\eqref{eq:action}: As $F_{ab}$ is antisymmetric, both the first and the second index pair of $Z^{abcd}$ can only contribute a completely antisymmetric part.
 	\begin{equation}
	 	Z^{(ab)cd} = Z^{ab(cd)} = 0.\label{eq:Zsym1}
 	\end{equation}
 	Therefore, each index pair has only $\nicefrac{4(4-1)}{2}=6$ degrees of freedom, which gives rise to the aforementioned possibility to rewrite it as $Z^{AB}$. 
 	
 	Furthermore, the action remains invariant under renaming the indices, providing
 	\begin{equation}
	 	Z^{abcd} = Z^{cdab},\label{eq:Zsym2}
 	\end{equation}
 	resulting in the total degrees of freedom of $\nicefrac{6(6+1)}{2}=21$.
 	
 	The discrepancy between the degrees of freedom of the conformal class of $\g$ (9 d.o.f.) and those of $Z$ (21 d.o.f.) clearly shows that some consistency conditions will have to exist and be fulfilled for $Z$ to be described by an effective metric $\g$ as in equation~\eqref{eq:Zeff}.
 	
 	\subsection{\texorpdfstring{The $6\times6$ representation of $Z$}{The 6x6 representation of Z}}\label{sec:6dim}
 	It is instructive to have a closer look at the representation of $Z$ as a symmetric $6\times6$ matrix, as indicated in equation~\eqref{eq:6x6} in the introduction and justified above. Written out, this matrix is
 	\begin{equation}
 	\kl{Z^{AB}}_{A,B\in\{1,\dots, 6\}} = \begin{pmatrix}
 	\epsilon & \zeta \\ \zeta^\dagger & \mu^{-1}
 	\end{pmatrix},\label{eq:Z6x6}
 	\end{equation}
 	where $\epsilon$ is the $3\times3$ permittivity matrix, $\mu^{-1}$ is the (inverse) $3\times3$ permeability matrix, and $\zeta$ is the $3\times3$ magneto-electric matrix. Here, $\epsilon$ and $\mi$ are real and symmetric, while $\zeta$ is real, but in general asymmetric. These link $\mathbf{E}, \mathbf{B}$ with $\mathbf{D}, \mathbf{H}$ in the following way\footnote{Just as the use of Franklin's \enquote{inconvenient} choice of the sign of the electric current (opposite to that of the flow of electrons) is a historical accident, so is the use of $\mi$ instead of $\mu$. We shall have to mention this again later on, as it sadly makes some subsequent results rather cumbersome in appearance.\label{fn:historylesson}}
 	\begin{alignat}{4}
	 	\mathbf{D} &=& \epsilon\; \mathbf{E}\; &+& \zeta\; \mathbf{B},\nonumber\\
	 	\mathbf{H} &=& \zeta^\dagger \mathbf{E}\; &+&\; \mu^{-1} \mathbf{B}.\label{eq:EBtoDH}
 	\end{alignat}
 	In terms of the $6\times6$-version of $Z$ this could be rewritten as
 	\begin{equation}
	 	\begin{pmatrix}
		 	\mathbf{D}\\\mathbf{H}
	 	\end{pmatrix} =
	 	\begin{pmatrix}
		 	\epsilon & \zeta \\ \zeta^\dagger & \mu^{-1}
	 	\end{pmatrix}
	 	\begin{pmatrix}
		 	\mathbf{E}\\\mathbf{B}
	 	\end{pmatrix}.
 	\end{equation}
 	This demonstrates the issue with this formalism for our purposes: All fields involved implicitly depend on the four-velocity $V^a$ of the observer. Therefore, the \enquote{constitutive matrices} mix in a quite messy way under Lorentz transformations (which are important in the flat space-time context), and even more so under general coordinate transformations (which become important, if we want to view $\g$ as an effective metric on a general, possibly curved background with physical metric $ g$). In the appendix we shall further investigate the relationship between $V^a$ and the constitutive matrices --- they will prove to be the elements of the Bel decomposition (also known as the orthogonal decomposition) with respect to given $V^a$.
 	
 	\subsection{Utilizing the conformal invariance}
 	As our counting of degrees of freedom showed, the \enquote{effective metric} is a conformal class of metrics rather than a metric as such. This in turn means that any representative of this class is equally valid, and thus we can simplify our analysis tremendously by focussing on the representative for which
 	\begin{equation}
	 	\det \g  = \det  g.
 	\end{equation}
 	Our constitutive tensor now takes on the form
 	\begin{equation}
	 	Z^{abcd} = \ed{2}\kl{\gi^{ac}\gi^{bd} - \gi^{ad}\gi^{bc}}.\label{eq:Zeffconffixed}
 	\end{equation}
 	If we use, for the time being, the effective metric $\g$ to raise and lower indices, it is then easy to show that
 	\begin{align}
	 	\kle{[\g]_{ae} \; [\g]_{bf}  Z^{efcd}}\; \kle{[\g]_{cm}\; [\g]_{dn} Z^{mnpq\vphantom{f}}}	\nonumber\\ =[\g]_{ae}\; [\g]_{bf}  Z^{efpq}
	 	 = \ed{2}\kl{\delta_a{}^p\delta_b{}^q -\delta_a{}^q\delta_b{}^p}.
 	\end{align}
 	This corresponds to the reciprocity or closure condition as found, for example, in \cite{HehlObukhov, RubilarPremetric}. Note that since we are not in a pre-metric setting it is unimportant to distinguish the two concepts.
 	
 	\section{Easing into the problem: \\
	A flat-space 3+1-decomposion}\label{sec:3+1}
 	While it is possible to immediately jump into the fully covariant, four-dimensional analysis, it is much more educational to first look at a more explicit 3+1 decomposition than in equation~\eqref{eq:Z6x6}. Furthermore, we will (for the time being) restrict attention to the flat space-time case, where $g = \eta = \mathrm{diag}(-1,1,1,1)$. The previous choice of a conformal factor turns to $\det \g = -1$.\footnote{Note that this differs from the choice in \cite{lrrAnalogue}, where the conformal invariance was used to set $\gi^{00} = -1$.} In the context of section \ref{sec:covariant}, this means that we consider going to Riemann normal coordinates. More specifically, we choose an observer with four-velocity $V = (1,0,0,0)^T$; spatial projection simply means limiting the range of an index to $\{1,2,3\}$, while time-projection is equivalent to setting the index equal to~0. This also means that all remaining indices are spatial and raised or lowered with a three-dimensional Kronecker symbol. Should we need four-dimensional indices, they will start from $a$, three-dimensional ones then from $i$.
 	It is easy to see that the definitions (see for example Appendix A in \cite{TrafoOptics2})
 	\begin{equation}
	 	\begin{gathered}
		 	\epsilon^{ij} =  -2 Z^{i0j0}; \qquad \mi^{ij} =  \ed{2}\;\eps^i{}_{kl}\,\eps^j{}_{mn}\,Z^{klmn};\\ 
		 	\zeta^{ij} = \eps^i{}_{kl}\,Z^{klj0}
	 	\end{gathered}\label{eq:defconstmatrices}
 	\end{equation}
 	satisfy equation~\eqref{eq:EBtoDH}.
 	
 	\subsection{\texorpdfstring{Vanishing magneto-electric $\zeta$}{Vanishing magneto-electric effects}}
 	A first step would now be to see what consistency conditions can be extracted under the simplifying assumption of a vanishing magneto-electric $\zeta$. Inserting equation~\eqref{eq:Zeffconffixed} into equations~\eqref{eq:defconstmatrices}, we find that
 	\begin{equation}
	 	0 \stackrel{!}{=}\zeta^{ij} = -  (\eps^i{}_{kl} \gi^{l0}) \gi^{kj}.\label{eq:3Dzeta0}
 	\end{equation}
 	From this it can be deduced that vanishing magneto-electric effects imply
 	\begin{equation}
	 	\gi^{i0} = 0.
 	\end{equation}
 	Using this, we get for the other two constitutive matrices:
 	\begin{align}
	 	\epsilon^{ij} &= \gi^{ij} \gi^{00};\\
	 	\mi^{ij} &= -{1\over2}\,
	 	\eps_{ikl}\,\eps_{jmn}
	 	\left( \gi^{km} \gi^{ln} \right).
 	\end{align}
 	Thus, $\g^{-1}$ block-diagonalizes. Since we know that $\det \g = -1$, we therefore can write this block structure as
 	\begin{equation}
	 	\renewcommand{\arraystretch}{1.25}
	 	\kl{\gi^{ab}}_{a,b\in \{ 0,\dots,3 \}} \ifed \begin{pmatrix}
		 	- \ed{\det \kl{\gamma^{ij}}} & 0\\ 0 & \gamma^{ij}
	 	\end{pmatrix}.
 	\end{equation}
 	Combining this with the following variant of Cramer's rule for $3\times3$ matrices,
 	\begin{equation}
	 	\eps_{ikl}\,\eps_{jmn} \{ X^{km} X^{ln} \} = 2 \det(X) \; X^{-1}_{ij},
 	\end{equation}
 	we can then reduce the equations for $\epsilon^{ij}$ and $\mi^{ij}$ to\footnote{Remember that spatial indices are raised and lowered with the three-dimensional Kronecker symbol.}
 	\begin{equation}
	 	\mu^{-1}_{ij}  =  \det(\gamma^{\circ\circ})\;  \gamma^{-1}_{ij},\qquad 	\Longleftrightarrow \qquad \mu^{ij} =  {\gamma^{ij}\over \det(\gamma^{\circ\circ})}\label{eq:mu3+1}
 	\end{equation}
 	and
 	\begin{equation}
	 	\epsilon^{ij} = {\gamma^{ij}\over \det(\gamma^{\circ\circ})} = \mu^{ij}.\label{eq:conscond3+1}
 	\end{equation}
 	This last equation, \eqref{eq:conscond3+1}, is exactly the consistency condition we were after. If it is fulfilled, we can write $\g^{-1}$ then as
 	\begin{align}
 	\gi^{ab} &= 
	\begin{pmatrix}
	 	-\sqrt{\det(\epsilon^{\circ\circ})} & 0 \\ 0 &  \displaystyle{\epsilon^{ij}\over \sqrt{\det(\epsilon^{\circ\circ})}}
 	\end{pmatrix}
 	\nonumber\\&= 
 	\begin{pmatrix}
	 	-\sqrt{\det(\mu^{\circ\circ})} & 0 \\ 0 &  \displaystyle{\mu^{ij}\over \sqrt{\det(\mu^{\circ\circ})}}
 	\end{pmatrix}.\label{eq:geff3+1zerozeta}
 	\end{align}
 	This particular result is well known and can, for example be found in \cite{TrafoOptics,AnalogueSurvey,TrafoOptics2}. Of course the matching condition $\epsilon^{ij}=\mu^{ij}$ does not hold for naturally occurring media.\footnote{Already a quick check on Wikipedia or in your favorite material data reference table will show this.} It is only with the development of modern meta-materials that the $\epsilon^{ij}=\mu^{ij}$ matching condition becomes plausible physics.
 	
 	To see what the effective metric (\emph{not} the inverse effective metric!) would be, one now needs to invert the matrix~\eqref{eq:geff3+1zerozeta}. Doing this, we simply arrive at our final results for zero magneto-electric effects:
\begin{equation}
[g_\mathrm{eff}]_{ab} = 
		 	\begin{pmatrix}
			 	-\det([\gamma]^{\circ\circ}) &  0\\ 
			 	0 & \gamma^{-1}_{ij}
		 	\end{pmatrix}.
\end{equation}	
This implies:	
 	\begin{subequations}
 		\begin{align}
		 	[g_\mathrm{eff}]_{ab}
		 	&= \begin{pmatrix}
		 	-\sqrt{\det(\mu^{\circ\circ})}^{-1} &  0\\ 
		 	0 & \sqrt{\det(\mu^{\circ\circ})}\;\mu^{-1}_{\,ij}
		 	\end{pmatrix},\\
		 	&=\begin{pmatrix}
		 	-\sqrt{\det(\epsilon^{\circ\circ})}^{-1} &  0\\ 
		 	0 & \sqrt{\det(\epsilon^{\circ\circ})}\;\epsilon^{-1}_{\,ij}
		 	\end{pmatrix}.
	 	\end{align}
 	\end{subequations}
 	
 	\subsection{\texorpdfstring{Non-vanishing magneto-electric $\zeta$}{Non-vanishing magneto-electric effects}}
 	The big difference, obviously, is that with non-vanishing magneto-electric effects equation~\eqref{eq:3Dzeta0} does not hold. This complicates the algebra --- but not in an impossible manner. Setting
 	\begin{equation}
 		\beta^{i} \defi \gi^{0i},
 	\end{equation}
 	and, again using the conformal freedom to set $\det \gi^{ab} = -1$, we consider the following, Kaluza--Klein-inspired form\footnote{As for the distinction between Kaluza--Klein and Arnowitt--Deser--Misner formulations, note that they are dual to each other: The same decomposition is applied either to the metric (ADM, see \cite{ADM}), or to the inverse metric (Kaluza--Klein, see \cite{Klein26}). For a modern textbook treatment, see chapter~X, appendices~6 through 9 of reference \cite{ZeeGR}. This ADM versus KK duality holds in the sense of the cotangent space being dual to the tangent space. This distinction is independent of additional considerations of dimensionality.} for $\gi$:
 	\begin{equation}
	 	\gi^{ab} =  \begin{pmatrix}
	 	-\det(\gamma^{-1}_{\circ\circ}) + \gamma^{-1}_{kl} \beta^k \beta^l & \beta^j\\ \beta^i & \gamma^{ij} 
	 	\end{pmatrix} .\label{eq:geffKaluzaKlein}
 	\end{equation}
 	Clearly, equation~\eqref{eq:mu3+1}, the result for $\mi$ from the previous calculation, remains the same. However, the equations for $\zeta$ and $\epsilon$ will change and become more difficult to deal with. It is useful to distinguish the earlier mentioned two ways to look at the consistency conditions: In the first case, one wants to take a given metric $\gi^{ab}$ and see with what material this metric could be achieved. After a bit of algebra (such as inverting $\gamma^{ij}$ as defined in equation~\eqref{eq:geffKaluzaKlein}), this can easily be done by looking at the following rewritten defining equations for the constituent matrices:
 	\begin{subequations}\label{eq:zeta3+1}
 		\begin{align}
	 		\epsilon^{ij} =&  
	 		\left(\gamma^{ij} \{\det(\gamma^{-1}_{\circ\circ}) - \gamma^{-1}_{kl} \beta^k \beta^l \}  + \beta^i \beta^j \right);\\
	 		\mu^{ij} =&  {\gamma^{ij}\over \det(\gamma^{\circ\circ})};\\
	 		\zeta^{ij} =& -{1\over2}\;
	 		\left(  \eps^{i}{}_{kl}\beta^l  \gamma^{kj}  \right).
 		\end{align}
 	\end{subequations}
	Should this set of equations not hold simultaneously, then the given metric \emph{cannot} be interpreted as an effective metric in macroscopic electrodynamics.
	
	The other way of looking at the consistency conditions is more involved and requires actually finding a concrete form of this condition. For this, take equation~\eqref{eq:zeta3+1} and use it to rewrite $\epsilon$ as
	\begin{equation}
	\epsilon^{ij} =  
	\mu^{ij}\, (1 - \mu^{-1}_{\,kl} \beta^k \beta^l )  + \beta^i \beta^j.\label{eq:conscond3+1zeta}
	\end{equation}
	This is the consistency condition we were looking for. Thus, if you are \emph{given} the optical properties ($\epsilon$, $\mu$, $\zeta$) --- \emph{and they fulfill this consistency condition} --- then you can calculate the effective metric via
	\begin{equation}
	\gamma^{ij}  = {\mu^{ij}\over \sqrt{\det(\mu^{\circ\circ})}};
	\end{equation}
	\begin{equation}
	\beta^m =   \sqrt{\det(\mu^{\circ\circ})} \; \eps^{mk}{}_i \; \mu^{-1}_{\,jk}\; \zeta^{ij};\label{eq:beta3+1}
	\end{equation}
	and insert in equation~\eqref{eq:geffKaluzaKlein} to arrive at:
	\begin{equation}
	[g_\mathrm{eff}^{-1}]^{ab} = \begin{pmatrix}
	-\sqrt{\det(\mu^{\circ\circ})}\; (1- \mu^{-1}_{\,kl} \beta^k \beta^l )& \beta^j\\ \beta^i &\displaystyle {\mu^{ij}\over\sqrt{\det(\mu^{\circ\circ})} }
	\end{pmatrix}.\label{eq:NonZeroKaluzaKlein}
	\end{equation}
	This could, in principle, be turned into an equivalent formula involving $\epsilon^{ij}$, but the results are not particularly edifying. In either case, if the consistency condition~\eqref{eq:conscond3+1zeta} is not satisfied, then the medium is simply \emph{not equivalent} to an effective metric.
	
	Doing either of these, we can then evaluate the effective metric $[g_{\mathrm{eff}}]_{ab}$ itself. In general, the inversion of the Kaluza--Klein decomposition~\eqref{eq:NonZeroKaluzaKlein} reads:
	\begin{widetext}
	\begin{equation}
		\renewcommand{\arraystretch}{1.25}
		[g_\mathrm{eff}]_{ab} = 
		\begin{pmatrix}
		-\det(\gamma^{\circ\circ}) &  \det(\gamma^{\circ\circ})\, \gamma^{-1}_{jk}\beta^k\\ 
		\det(\gamma^{\circ\circ})\,\gamma^{-1}_{ik}\beta^k 
		&\;\;\gamma^{-1}_{ij} - \det(\gamma^{\circ\circ})( \gamma^{-1}_{ik} \beta^k) ( \gamma^{-1}_{jl}\beta^l)
		\end{pmatrix}.
	\end{equation}
	Inserting the consistency condition~\eqref{eq:conscond3+1zeta} we arrive at
	\begin{subequations}
		\begin{align}
			[g_\mathrm{eff}]_{ab} &= 
			\begin{pmatrix}
			-\sqrt{\det(\mu^{\circ\circ})}^{-1} &  \mu^{-1}_{\,jk}\beta^k\\ 
			\mu^{-1}_{\,ik}\beta^k 
			&\sqrt{\det(\mu^{\circ\circ})}\kl{\mu^{-1}_{\,ij} - ( \mu^{-1}_{\,ik} \beta^k) ( \mu^{-1}_{\,jl}\beta^l)}
			\end{pmatrix},\\
			&= \begin{pmatrix}
			- \sqrt{\det \epsilon^{\circ\circ}}^{-1}(1-\epsilon_{kl}^{-1}\beta^k\beta^l) & \epsilon_{jk}^{-1}\beta^k\\
			\epsilon^{-1}_{ik}\beta^k & \sqrt{\det\epsilon^{\circ\circ}}(\epsilon^{-1}_{ij})
			\end{pmatrix}, \qquad \text{with~} \beta^m \mathrel{\mathop:}= \sqrt{\det(\epsilon^{\circ\circ})} \; \varepsilon^{mk}{}_i \; \epsilon^{-1}_{\,jk}\; \zeta^{ij}.
		\end{align}
	\end{subequations}
	\end{widetext}
 	
 	\section{Generalizing to a fully covariant approach}\label{sec:covariant}
 	The general idea for upgrading the analysis to a fully covariant approach is that the analysis in Minkowski space-time can be seen as the case of an arbitrary space-time in Riemann normal coordinates. Remember that we can use the temporal and spatial projection operators, respectively $t^{a}{}_b = -V^aV_b$ and $h^a{}_b = g^{a}{}_b + V^a V_b$, to write any vector as
 	\begin{equation}
 	X^a = \delta^{a}{}_b X^b = t^a{}_b X^b + h^a{}_b X^b.
 	\end{equation}
 	This naturally and obviously extends to higher-degree tensors. Also, note the signs due to Lorentz signature. Then, effectively, in our earlier calculation spatial indices $i = 1, 2, 3$ correspond to spatially-projected indices and time-like indices (indices set to zero) correspond to a contraction with the given four-velocity $V$.\footnote{Strictly speaking, the index should be hit with the temporal projector $t^a{}_b$ --- but the actual information contained in these processes is the same.} Any three-dimensional Kronecker symbol $\eps^{ijk}$ corresponds then to a contraction of the four-dimensional one with the four-velocity. Summarizing, we get the following set of translation rules:
 	\begin{subequations}
 		\begin{align}
		 	X^i &\longrightarrow h^a{}_b X^b\\
		 	X^0 &\longrightarrow V_a X^a\qquad \Longleftrightarrow\qquad X^0 \longrightarrow t^a{}_b X^b\\
		 	\eps^{ijk} &\longrightarrow \eps^{abcd}V_a.
	 	\end{align}
 	\end{subequations}
 	A quick consistency check: If we were to use these translation rules on the definition of the constitutive matrices~\eqref{eq:defconstmatrices}, we arrive at just the equations~\eqref{eq:WandDuals} in terms of the constitutive matrices $\epsilon$, $\mi$ and $\zeta$:
 	\begin{subequations}\label{eq:CovConstMatrices}
 		\begin{align}
	 		\epsilon^{ab} &\defi - 2 Z^{acbd} V_c V_d, \label{eq:defeps4}\\
	 		\mi^{ab} &\defi \frac{1}{2} \eps^{ca}{}_{ef}\eps^{db}{}_{gh} Z^{efgh} V_c V_d, \label{eq:defmu4}\\
	 		\zeta^{ab} &\defi \eps^{ca}{}_{ef} Z^{efbd} V_c V_d. \label{eq:defZeta4}
 		\end{align}
 	\end{subequations}
 	This links the previously considered special case with the general orthogonal decomposition presented in the appendix. Inserting the mimicking conditions~\eqref{eq:Zeff}, we get:
 	\begin{subequations}
 		\begin{align}
	 		\epsilon^{ab} &=& -&\left( \gi^{ab}\;\gi^{cd} - \gi^{ac}\; \gi^{bd} \right) V_c V_d;\\ 
	 		\mu^{-1}_{ab} &=&& \eps_{aefc}\,\eps_{bmnd}\left( \gi^{em}\; \gi^{fn} \right)   V^c V^d; \\
	 		\zeta_a{}^b&=& -&  (\eps_{amnd} \gi^{mc})\;  \gi^{nb} \; V_c V^d .
 		\end{align}
 	\end{subequations}	
 	
 	However, there are two ingredients missing: Looking back at our equations in section \ref{sec:3+1} we note that we frequently encounter both the inverses of $3\times3$ matrices and their determinants. Both notions are not as straightforwardly translated. To solve this, we shall use the Moore--Penrose pseudo-inverse $A^\#$ (see, e.g., \cite{MooreInverse,PenroseInverse,GenInverses}\footnote{\cite{GenInverses} also contains some more historic references about other (re)discoveries of the pseudo-inverse.}) and the pseudo-determinant\footnote{Early notions of the pseudo-determinant can be found in \cite{KhatriPDET}, while more modern appearances include \cite{KnillPDET,etalPDET}. Written as $\det' (A)$, a similar notion for operators can be found in the quantum field theory literature in \cite{AspectsSym} and probably even earlier. This notation has been adopted, for example, in \cite{BivsBi}.} $\pdet (A)$, defined for a general square $n\times n$ matrix $A$ with eigenvalues $\lambda_i$ as follows:
 	\begin{equation}
 		\pdet (A) = \prod_{\substack{i=1 \\\lambda_i \neq 0}}^{\mathrm{rank}(A)}\lambda_i.
 	\end{equation}
 	Furthermore, the following identities hold for the pseudo-determinant, with the last equality valid for (anti-)\break symmetric or (anti-)Hermitian matrices:\footnote{For general (asymmetric) matrices, this can be generalized to \begin{equation*}\pdet(A\;A^\dagger) = \det\kl{\kle{\mathds{1}-A \; A^\#} + A\;A^\dagger}\end{equation*} using the singular value decomposition of $A$.}
 	\begin{subequations}
 		\begin{align}
 			\det\kl{\mathds{1}+zA} &= \pdet (A)\; z^{\mathrm{rank}(A)} + \mathcal{O}\kl{z^{\mathrm{rank}(A)-1}},\\
			\pdet (A) &= \lim_{z\to 0} \frac{\det\kl{A+z\mathds{1}}}{z^{n-\mathrm{rank}(A)}} = \lim_{z\to 0} \frac{\det\kl{A+z\mathds{1}}}{z^{\mathrm{nullity}(A)}},\\
					&= \det \kl{\kle{\mathds{1} - A\; A^\#} + A}.		
 		\end{align}
 	\end{subequations}
 	
 	Then note that while the generally covariant $\mi$ and $\epsilon$ remain symmetric, due to their orthogonality to $V^a$ they will not have full rank as $4\times4$ matrices. Put differently, the null-space of $\epsilon$ or $\mi$ is one-dimensional, any two projection operators onto this null-space therefore proportional to each other. As $V^a V^b = - t^{ab}$ is a projector onto this null-space of $\mi$ and $\epsilon$, this has to be proportional to the corresponding $\kle{\mathds{1} - A\; A^\#}^{ab}$. Note that $\kle{t^{\bullet\bullet}}^\#_{ab} = t_{ab} = -V_a V_b$. Furthermore, as we want the $3+1$ case to drop out if we chose $V = (1,0,0,0)^T$, we can see that
 	\begin{equation}
 		\kle{\mathds{1} - A\; A^\#}^{ab} = - t^{ab} = +V^a V^b.
 	\end{equation}
 	Put to use on the pseudo-determinant, we can then give it in terms of a perfectly well-behaved, standard determinant:
 	\begin{equation}
 		\pdet (\epsilon^{ab}) = \det \kl{-t^{ab} + \epsilon^{ab}} = \det \kl{V^aV^b + \epsilon^{ab}}
 	\end{equation}
 	
 	Now we are in the position to actually generalize the $\det \epsilon^{ij}$ or $\det \mi^{ij}$ terms to a fully covariant formalism that appear in, for example, equation~\eqref{eq:geff3+1zerozeta} or \eqref{eq:beta3+1}. As determinants of a tensor pick up determinants of the physical metric under general coordinate transformations, we need the following rules for promoting determinants to quantities that behave as scalars under general coordinate transformations:\footnote{Note that as we only take determinants of symmetric matrices, $S^{[ab]}=0$, the bullet notation we employ is sufficient. This means, in terms of translation rules, that
 		\begin{equation*}
 		S_{\circ\circ}\quad \longrightarrow \quad S_{\bullet\bullet},\qquad 		S^{\circ\circ} \quad \longrightarrow \quad S^{\bullet\bullet}.
 		\end{equation*}}
 	\begin{equation}
 		\det (A^{ij}) \qquad \longrightarrow \qquad \frac{\pdet (A^{ab})}{-\det (g^{ab})}.
 	\end{equation}
 	 	
 	We summarized all important rules in table~\ref{tab:translation}.
 	
 	\begin{table}
 		\centering
	 	\begin{tcolorbox}[width=.75\linewidth]
	 		\renewcommand{\arraystretch}{1.25}
		 	\centering
	 		\begin{tabular}{rcl}
	 			$X^i$ &$\qquad \longrightarrow\qquad $& $h^a{}_b X^b$\\
	 			$X^0$ & $\qquad\longrightarrow\qquad$ &$V_a X^a$\\
	 			or $X^0$ &$\qquad\longrightarrow\qquad$& $t^a{}_b X^b$\\
	 			$\eps^{ijk}$ &$\qquad\longrightarrow\qquad$  &$\eps^{abcd}V_a$\\
	 			$(A^{ij})^{-1}$ &$\qquad\longrightarrow\qquad$&$(A^{ab})^\#$\\
	 			$\det (S^{\circ\circ})$ &$\qquad \longrightarrow \qquad$& $\frac{\pdet (S^{\bullet\bullet})}{-\det (g^{\bullet\bullet})}$
	 		\end{tabular}
	 	\end{tcolorbox}
	 		 		\caption{Translating 3+1 terms to fully covariant terms.}
	 		 		\label{tab:translation}
 	\end{table}
 	
 	\subsection{Zero magneto-electric effects}\label{sec:covzero}
 	Again, vanishing magneto-electric effects will greatly expedite the calculation. And as we shall see later in subsection~\ref{sec:surprise}, this now is more than just a pedagogical introduction --- it actually has a connection to the final form of the consistency condition. With our translation rules in place, we can immediately proceed and get for the expression for the inverse of the effective metric
 	\begin{equation}
	 	\gi^{ab} = - \sqrt{\frac{\pdet(\epsilon^{\bullet\bullet})}{-\det(g^{\bullet\bullet})}}\; V^a V^b +  \sqrt{-\det(g^{\bullet\bullet})\over{\pdet(\epsilon^{\bullet\bullet})}}\; \epsilon^{ab},
 	\end{equation}
 	while our consistency condition is turned into
 	\begin{equation}
	 	\epsilon^{ab} = \kle{\mi_{\bullet\bullet}^\#}^{ab}.
 	\end{equation}
 	If we then were to define
 	\begin{eqnarray}
	 	\mu^{ab} \defi \kle{\kle{\mu^{-1}_{\bullet\bullet}}^\#}^{ab},
 	\end{eqnarray}
 	we could simplify this to the familiar
 	\begin{equation}
	 	\epsilon^{ab} = \mu^{ab}.
 	\end{equation}
 	However, the hidden mix of inverse (from the traditional notation $\mi$ to link magnetic field to excitation, unlike for the permittivity) and pseudo-inverse has to be kept in mind. Again, this is related to the historical artefact of the naming of $\mi$, as mentioned in footnote~\ref{fn:historylesson}.
 	The effective metric itself now takes on any of the following forms:
	\begin{subequations}
	 	\begin{align}
		 	\kl{g_\mathrm{eff}}_{ab} &= \frac{\pdet\kl{\gamma^{\bullet\bullet}}}{-\det \kl{g^{\bullet\bullet}}}t_{ab} + \kle{\gamma^{\bullet\bullet}}^\#_{ab},\\
		 	&= -\sqrt{\frac{-\det(g^{\bullet\bullet})}{\pdet(\epsilon^{\bullet\bullet})}}\, V_a V_b + \sqrt{\frac{\pdet(\epsilon^{\bullet\bullet})}{-\det(g^{\bullet\bullet})}} [\epsilon^{\bullet\bullet}]^\#_{ab},\\
		 	&= -\sqrt{\frac{-\det(g^{\bullet\bullet})}{\pdet(\mu^{\bullet\bullet})}}\, V_a V_b + \sqrt{\frac{\pdet(\mu^{\bullet\bullet})}{-\det(g^{\bullet\bullet})}} \mu^{-1}_{ab}.
	 	\end{align}
	\end{subequations}
 	
 	\subsection{Non-zero magneto-electric effects}
 	The starting point here are now the consistency conditions~\eqref{eq:conscond3+1zeta}, the $0i$ components of the metric~\eqref{eq:beta3+1}, together with the result for the Kaluza--Klein decomposition~\eqref{eq:NonZeroKaluzaKlein}. All of these are turned into the corresponding, fully covariant versions by straightforwardly applying the previously derived rules.
 	
 	First, take a look at what happens to the three-vector $\beta^i$:
 	\begin{equation}
	 	\beta^i \qquad \longrightarrow \qquad \beta^e  =\sqrt{ \pdet(\mu^{\bullet\bullet})\over{-\det(g^{\bullet\bullet})  }} \; \eps^{ecad} \; \mu^{-1}_{\,bc}\; \zeta_a{}^{b}  V_d.
 	\end{equation}
 	We can immediately see that the four-vector $\beta^e$ satisfies
 	\begin{equation}
	 	\beta^e\, V_e = 0,
 	\end{equation}
 	a transversality result we can immediately put to use to see that
 	\begin{equation}
	 	\mu^{-1}_{\,ij}\beta^i \beta^j \qquad \longrightarrow \qquad \mu^{-1}_{\,ab}\beta^a\beta^b.
 	\end{equation}
 	From this we can derive the inverse effective metric\footnote{Had we chosen to turn $\beta^i$ into the equivalent tensorial form \smash{$\tilde{\beta}^{ab} \defi t^{a}{}_c [g_\text{eff}^{-1}]^{cd} h_d{}^b$}, the transversality would have been $\tilde{\beta}^{ab}V_b = 0$, and the combination $\beta^a V^b$ would be equal to $\tilde{\beta}^{ba}$.}:
 	\begin{align}
 	\gi^{ab} =& - \sqrt{{\pdet(\mu^{\bullet\bullet})}\over{-\det(g^{\bullet\bullet})}}  \left( 1-  \mu^{-1}_{cd} \beta^c \beta^d\right)\; V^a V^b \nonumber\\&+  
 	V^a \beta^b + \beta^a V^b + 
 	\sqrt{-\det(g^{\bullet\bullet})\over{\pdet(\mu^{\bullet\bullet})}}\; \mu^{ab}.
 	\end{align}
 	
 	The consistency condition is simply turned into the fully Lorentz-invariant, covariant equation
 	\begin{equation}
	 	\epsilon^{ab} = \mu^{ab} (1 - \mu^{-1}_{\,cd} \beta^c \beta^d )  + \beta^a \beta^b. 
 	\end{equation}
 	
 	For the effective metric itself, we can use the fact that $\gamma$ and $\gamma^\#$ will again be orthogonal to $V$. The somewhat long expression we get is\vspace{3\baselineskip}
 	\begin{widetext}
 	\begin{subequations}
 		\begin{equation}
 		\left[g_\text{eff}\right]_{ab} = \frac{-\det \left(g_{\,\bullet\bullet}\right)}{\mathrm{pdet}\left(\left[\gamma^{\bullet\bullet}\right]^\#_{\bullet\bullet}\right)} \left( t_{ab} - V_{a}\left[\gamma^{\bullet\bullet}\right]^\#_{bc}\beta^c - V_{b}\left[\gamma^{\bullet\bullet}\right]^\#_{ac}\beta^c + \frac{\mathrm{pdet}\left(\left[\gamma^{\bullet\bullet}\right]^\#_{\bullet\bullet}\right)}{-\det \left(g_{\,\bullet\bullet}\right)} \left[\gamma^{\bullet\bullet}\right]^\#_{ab} - \left[\gamma^{\bullet\bullet}\right]^\#_{ac}\beta^c \left[\gamma^{\bullet\bullet}\right]^\#_{bd}\beta^d\right).
 		\end{equation}
 		More specifically, in terms of $\mu$,
 		\begin{equation}
 		\left[g_\text{eff}\right]_{ab} = \sqrt{\frac{\mathrm{pdet}\left[\left[\mu^{-1}\right]^{\bullet\bullet}\right]}{-\det \left(g^{\bullet\bullet}\right)}} t_{ab} - \left( V_{a} \,\mu^{-1}_{\,bc}\,\beta^c + V_{b}\,\mu^{-1}_{\,ac}\,\beta^c\right)  + \sqrt{\frac{-\det \left(g^{\bullet\bullet}\right)}{\mathrm{pdet}\left(\left[\mu^{-1}\right]^{\bullet\bullet}\right)}}\left(\mu^{-1}_{\,ab} - \,\mu^{-1}_{\,ac}\,\beta^c \,\mu^{-1}_{\,bd}\beta^d\right).
 		\end{equation}
 		Alternatively, we can also write this in terms of $\epsilon$ as
 		\begin{align}
 		\left[g_\text{eff}\right]_{ab} = - \sqrt{\frac{-\det g^{\bullet\bullet}}{\mathrm{pdet}\epsilon^{\bullet\bullet}}}\left(1-\epsilon_{cd}^\#\beta^c\beta^d\right) V_aV_b - V_a \epsilon_{bd}^\# \beta^d - V_b \epsilon_{ad}^\# \beta^d + \sqrt{\frac{\mathrm{pdet}\epsilon^{\bullet\bullet}}{-\det g^{\bullet\bullet}}} \epsilon^\#_{ab},
 		\end{align}
 		where now
 		\begin{equation}
 		\beta^e  =\sqrt{ \mathrm{pdet}(\epsilon^{\bullet\bullet})\over{-\det(g^{\bullet\bullet})  }} \; \varepsilon^{ecad} \; \epsilon^{\#}_{\,bc}\; \zeta_a{}^{b}  V_d.
 		\end{equation}
 	\end{subequations}
 	\end{widetext}
 	
 	\subsection{A new look at the consistency condition}\label{sec:surprise}
 	On physical grounds, the \enquote{light-cones} of $g_\mathrm{eff}$ will have to lie inside the light-cones of the physical metric $g$. Therefore, for any \emph{physical} four-velocity $U^a$, the quantity
 	\begin{equation}
	 	Q= [g_\mathrm{eff}]_{ab} U^a U^b
 	\end{equation}
 	will be negative. Now look for the minimum of $Q$ by solving the Lagrange multiplier problem
 	\begin{equation}
	 	L = [g_\mathrm{eff}]_{ab} \,U^a U^b - \lambda (g_{ab}\, U^a U^b +1),
 	\end{equation}
 	and call this minimum $V$. Now adopting Riemann normal coordinates ($g_{ab}\to \eta_{ab}$) and going to the rest-frame of $V$ (so $V^a\to(1,0,0,0)$) we can block-diagonalize the effective metric
 	\begin{equation}
	 	\kle{g_\mathrm{eff}}_{ab} = \begin{pmatrix}
	 	-\lambda & 0 \\ 0 & \kle{g_\mathrm{eff}}_{ij}
	 	\end{pmatrix}
 	\end{equation}
 	with inverse
 	\begin{equation}
	 	\kle{g_\mathrm{eff}}^{ab} = \begin{pmatrix}
	 	-\ed{\lambda} & 0 \\ 0 & \kle{g^{-1}_\mathrm{eff}}^{ij}
	 	\end{pmatrix}.
 	\end{equation}
 	
 	In particular, this means that, for this effective metric, there exists a rest-frame for an observer with four-velocity $V$ such that \emph{in this rest-frame the magneto-electric effects vanish}. Now this means that we can, for this specific observer(!), use the much simpler analysis of section \ref{sec:covzero}! Let us therefore call this the \emph{natural rest-frame} of the given medium.
 	
 	Thus, another possible approach to the problem is this: Assume we have found this $V$ for our given effective metric. We then define the corresponding permittivity as $\epsilon_V$ and the corresponding permeability as $\mu^{-1}_V$. What, then, would be the constitutive matrices $\epsilon$, $\mi$, and $\zeta$ of another observer with four-velocity $W$ in terms of these $\epsilon_V$ and $\mu^{-1}_V$?
 	
 	\subsection{Natural reference frame versus arbitrary observer}\label{sec:frames}
 	In order to answer the question at the end of the last subsection, let us first establish helpful notation for this. Choose any four-velocity $V^a$ and an arbitrary, not necessarily symmetric matrix $q^{ab}$ four-orthogonal to it:
 	\begin{equation}
	 	q^{ab}V_b = V_b \, q^{ba} = 0.
 	\end{equation}
 	Let us then define the following fourth-rank tensor
 	\begin{equation}
	 	 Q^{abcd} \defi V^a V^d  q^{bc} + V^b V^c  q^{ad} - V^b V^d  q^{ac} - V^a V^c  q^{bd}.
 	\end{equation}
 	Furthermore, let us use this tensor $ Q^{abcd}$ to define four more tensors by setting $q$ equal to one of the four \enquote{constitutive matrices} $\epsilon_V^{ab}$, $[\mu^{-1}_V]^{ab}$, $\zeta_V^{ab}$, and its transpose $[\zeta^T_V]^{ab}$ as measured with respect to four-velocity $V^a$:
 	
 	\begin{center}
 		\begin{subequations}
	 		\begin{minipage}{.45\textwidth}
	 			\begin{align}
	 			E^{abcd}_V &\defi  Q^{abcd}_{ q\rightarrow\epsilon_V},\hphantom{bladiebla}\\
	 			\addtocounter{equation}{+1}A^{abcd}_V &\defi  Q^{abcd}_{ q\rightarrow\zeta_V},
	 			\end{align}
	 		\end{minipage}\hfill
	 		\begin{minipage}{.45\textwidth}
	 			\begin{align}
		 			\addtocounter{equation}{-2}M^{abcd}_V &\defi  Q^{abcd}_{ q\rightarrow\mu^{-1}_V},\hphantom{bladiebla}\\
		 			\addtocounter{equation}{1}(A^T_V)^{abcd} &\defi  Q^{abcd}_{ q\rightarrow\zeta^{T}_V}.
	 			\end{align}
	 		\end{minipage}
	 	\end{subequations}
 	\end{center}
 
 	If we now compare this with the Bel-decomposed expression for the constitutive tensor $Z^{abcd}$ in equation~\eqref{eq:ZBel2}, we see that we can rewrite equation~\eqref{eq:ZBel2} in terms of these four tensors in the following way:
 	\begin{equation}
	 	Z^{abcd} = \ed{2}\kl{E_V + (\ast M_V\ast) + (\ast A_V) + (A^T_V \ast)}^{abcd}.
 	\end{equation}
 	While the right-hand side is implicitly dependent on the previously chosen four-velocity $V^a$, the left-hand side is general and independent of it. This, then, enables us to give deceptively simple expressions for how to calculate the \enquote{constitutive matrices} $\epsilon^{ab}_W$, $[\mu^{-1}_W]^{ab}$, and $\zeta_W^{ab}$ as seen by a different observer with four-velocity $W^a$ 
 	
 	\begin{flushright}
 		\begin{subequations}\label{eq:consmatW}
 			\begin{minipage}{.45\textwidth}
 				\begin{align}
 				\epsilon_W^{ab} &= -2 Z^{dacb} W_d W_c,\hphantom{blad}\label{eq:epsW}\\
 				\addtocounter{equation}{+1}\zeta_W^{ab} &= 2 (\ast Z)^{dacb} W_d W_c,\label{eq:zetaW}
 				\end{align}
 			\end{minipage}\hfill
 			\begin{minipage}{.45\textwidth}
 				\begin{align}
 				\addtocounter{equation}{-2}[\mu^{-1}_W]^{ab} &= 2(\ast Z \ast)^{dacb} W_d W_c,\hphantom{bla}\label{eq:muW}\\
 				\addtocounter{equation}{1}  \zt^{ab} &= 2(Z \ast)^{dacb} W_d W_c.
 				\end{align}
 			\end{minipage}
 		\end{subequations}
 	\end{flushright}
 
 	While it would be certainly possible to now give $\epsilon^{ab}_W$, $[\mu^{-1}_W]^{ab}$, and $\zeta_W^{ab}$ in full generality in terms of $\epsilon_V^{ab}$, $[\mu^{-1}_V]^{ab}$, $\zeta_V^{ab}$, $V^a$, and $W^a$, the resulting expressions would be stigmatized by being unilluminatingly and excessively complicated. Nevertheless, in special cases this will be much less of a problem. Also, the existence of closed-form expressions will prove useful when working numerically in this formalism.
 	Nonetheless, in appendix~\ref{sec:iso} we shall give an explicit example on how to use this. Specifically, we shall look at an isotropic medium in motion and regain the well known magneto-electric effect of moving media \cite{TrafoOptics2,ODell,Post}, of which the Fresnel--Fizeau effect is a special case\cite{Jackson,LanLifVIII}.
 	
 	\section{Conclusion}
 	In conclusion, we have seen that even going to generally covariant formulations of an effective metric given by macroscopic electrodynamics gives no additional physical results: There will always be a natural reference frame for a given medium such that in this frame the consistency condition reduces to the well-known result
 	\begin{equation}
	 	\epsilon^{ij} = \mu^{ij}.
 	\end{equation}
 	It remains to be seen how far-reaching or maybe even limiting this result proves to be. On physical grounds, however, this should not come as a surprise: The very nature of the effective metric $\g$ is to describe the given physics via a light cone --- in particular, this implies locality. And in classical electrodynamics it is well known (see \cite{Post}) that only non-local (and so, when Lorentz-transformed, non-instantaneous), or dissipative phenomena can give rise to non-vanishing magneto-electric $\zeta$. While both non-locality (through, for example, helical molecules) and dissipation (through electrical resistance in a medium)  \emph{are} obviously important effects in macroscopic electrodynamics, their effects will lead beyond mere Lorentzian geometries in an analogue model. It is useful to compare this with the physical arguments behind requiring a vanishing birefringence in the context of pre-metric electrodynamics, as done in reference \cite{HehlLaemmerzahl}.
 	
 	The covariant formulation we employed, however, should provide --- in the right context --- a great boon to presentations of macroscopic electrodynamics. In particular the concept of pseudo-inverses and pseudo-determinants provides a quite intuitive (and so far under-appreciated) mathematical technology. Therefore, it will prove useful to further disseminate this framework: When communicating with researchers with a background in relativity (who are used to treating microscopic electrodynamics fully covariantly), the different 3+1 notation inherited from the electrodynamics community, and the focus there on three-dimensional quantities, often complicates discussion. Vice versa, the fully covariant formulation can be used to make the covariant approach itself more appealing to people used to the three-dimensional quantities $\mathbf{E}$, $\mathbf{B}$, $\mathbf{D}$, and $\mathbf{H}$ on the one hand, and the corresponding $3\times 3$ matrices for permittivity $\epsilon$, (inverse) permeability $\mi$, and magneto-electric effects $\zeta$ (or their $6\times6$ matrix analogue as in equation~\eqref{eq:6x6}). Especially in the context of analogue space-times implemented via macroscopic electrodynamics, this translational device should prove helpful, as it is here that both respective communities have to come together.
 	
 	Note that nothing could prevent us from using a covariant polarization tensor $P^{ab}$ instead of the excitation tensor $G^{ab}$, thus generalizing the present discussion somewhat. However, this could not give rise to new physical insights and would rather only make the notation even more cumbersome in this particular context. Similarly, while the constitutive tensor in macroscopic electrodynamics is often immediately made complex-valued to deal with dissipation and dispersion, in the present context this runs into problems early on --- one would have to provide a physical interpretation of a complexified effective metric. While this might prove important for applications of electrodynamic, analogue space-times, it is far from obvious how to solve this problem.

 	\appendix
 	\section*{Acknowledgements}
 	This research was supported by the Marsden Fund, through a grant administered by the Royal Society of New Zealand. S.S. is also supported via a Victoria University of Wellington PhD scholarship. S.S. would like to thank Natalie Deruelle, Chris Fewster, Friedrich Hehl, and Dennis Rätzel for helpful discussions.
 	
 	\section{The Bel decomposition of the constitutive tensor}
 	
 	The Bel decomposition was originally developed as the orthogonal decomposition, with respect to a given four-velocity $V^a$, of the Riemann curvature tensor (see for example \cite{BelSem,Bel1,MatteDecom,BelTrans,GravEMagAna,DynLawSuEGR} and references therein; for unnamed appearances in the present context see for example \cite{BalNi}). In order to see how this comes about, it is useful to remind oneself of the orthogonal decomposition w.r.t. to an observer of four-velocity $V^a$ of some two-form, e.g. the electromagnetic field-strength tensor $F_{ab}$ or the excitation tensor $G_{ab}$:
 	
 	For all four-velocities $V^a$ there exist two uniquely determined vector fields $E^a$ and $B^a$, such that
 	\begin{equation}
	 	F_{ab} = V_a E_b - V_b E_a + \eps_{abcd}V^c B^d.\label{eq:Fdecomp}
 	\end{equation}
 	A proof can be found in \cite{GourSR}, page~83ff; see also page~493 therein. 
 	
 	This enables us to have a rigorous look at section~\ref{sec:6dim}: Together with the symmetries of $Z$ given in equations~\eqref{eq:Zsym1} and \eqref{eq:Zsym2}, we can then deduce that there exist $Y_A$ (a collection of two-forms labelled by $A$) and a symmetric $6\times6$ matrix $X^{AB}$, such that
 	\begin{equation}
	 	Z^{abcd} = Y_A^{ab} X^{AB} Y_B^{cd}.\label{eq:ZeqYXY}
 	\end{equation}
 	Each of the six $Y_A^{ab}$ decomposes as the field-strength tensor for a given four-velocity $V^a$ with corresponding vector fields $E_A$ and $B_A$.\footnote{The naming is chosen such that their role in the corresponding version of equation~\eqref{eq:Fdecomp} is clear; this is not to mean that they are six electric or magnetic fields!} Inserting these decompositions in equation~\eqref{eq:ZeqYXY} and collecting terms, we can define three separate matrices from four separate terms:
 	\begin{align}
	 	 W_\epsilon^{ab} &\defi E_A^a X^{AB} E_B^b& W_\zeta^{ab} &\defi E_A^a X^{AB} B^b_B\\
	 	[W_\zeta^T]^{ab} &\defi B_A^a X^{AB} E_B^b&   W_\mu^{ab} &\defi B_A^a X^{AB} B^b_B
 	\end{align}
 	With these definitions, $Z$ decomposes in the following manner:
 	\begin{align}
	 	Z^{abcd} =& V^b V^d W_\epsilon^{ac} + V^a V^c W_\epsilon^{bd} - V^a V^d W_\epsilon^{bc} - V^b V^c W_\epsilon^{ad}\nonumber\\
	 	 &\;+V^f (\eps^{ab}{}_{ef} W_\mu^{eg} \eps^{cd}{}_{gh}) V^h\nonumber\\
	 	&\; +\kl{W_\zeta^{ag} V^b - W_\zeta^{bg} V^a} \eps^{cd}{}_{gh} V^h\nonumber\\
	 	& \;+V^f \eps^{ab}{}_{ef} \kl{[W_\zeta^T]^{ec} V^d - [W_\zeta^T]^{ed} V^c}.\label{eq:ZBel1}	 	
 	\end{align}
 	Now define
 	\begin{subequations}\label{eq:WtoConstMat}
		\begin{align}
		 	-2 W_\epsilon &= \epsilon,\\
		 	2 W_\zeta &= \zeta,\\
		 	2 W_\mu &= \mi.
	 	\end{align}
 	\end{subequations}
 	It is noteworthy that the above procedure bears a close relationship to the left, right and double-dual as usually defined for the Riemann tensor, see e.g. \cite{MTW}, as
 	\begin{subequations}\label{eq:WandDuals}
 		\begin{align}
	 		W_\epsilon^{bd} &= V_a V_c Z^{abcd},\\
	 		W_\mu^{bd} &= V_a V_c \kl{\ast Z \ast}^{abcd},\\
	 		[W_\zeta^T]^{bd} &= V_a V_c \kl{\ast Z}^{abcd},\\
	 		W_\zeta^{bd} &= V_a V_c \kl{Z\ast}^{abcd}.
 		\end{align}
 	\end{subequations}
 	After some longer index algebra, equation~\eqref{eq:ZBel1} can be turned into
 	\begin{widetext}
 	\begin{align}
	 	Z^{abcd} = &\ed{2}\kl{V^a V^d \epsilon^{bc} + V^b V^c \epsilon^{ad} -V^b V^d \epsilon^{ac} - V^a V^c \epsilon^{bd} }\nonumber\\
	 	&+ \ed{8} \eps^{ab}{}_{ef} \eps^{cd}{}_{gh}\kl{V^f \mi^{eg} V^h + V^e \mi^{fh} V^g - V^e \mi^{fg} V^h - V^f \mi^{eh} V^g}\nonumber\\
	 	&+ \ed{4} \eps^{ab}{}_{ef} \kl{\zeta^{fc}V^d V^e + \zeta^{ed} V^c V^f - \zeta^{ec}V^d V^f - \zeta^{fd} V^c V^e}\nonumber\\
	 	&+ \ed{4} \eps^{cd}{}_{gh} \kl{\zt^{bg} V^a V^h  + \zt^{ah} V^b V^g   - \zt^{ag} V^b V^h  - \zt^{bh} V^a V^g }.\label{eq:ZBel2}
 	\end{align}
 	To get another way of writing this decomposition, make use of the spatial projection $h^{ab} \defi g^{ab} + V^a V^b$ and the time-projection $t^{ab} \defi -V^a V^b$. Noting that
 	\begin{equation}
	 	\epsilon_{cd}{}^{af} \epsilon^{ebcd} V_e V_f = -2 (g^{ab}+V^a V^b) = -2h^{ab},\label{eq:projection}
 	\end{equation}
 	and using
 	\begin{equation}
	 	g^{b_1 c_1}\cdots g^{b_n c_n} \eps_{c_1\dots c_n} \eps^{a_1\dots a_n} = -n! g^{b_1 c_1}\cdots g^{b_n c_n} \delta^{a_1}{}_{[c_1}\cdots\delta^{a_n}{}_{c_n]},
 	\end{equation}
 	explicitly written out for $n=4$, one gets
 	\begin{align}
	 	Z^{abcd} = &\frac{1}{2} \left( V^d V^a \epsilon^{bc} -V^c V^a \epsilon^{bd} + V^c V^b \epsilon^{ad} - V^d V^b \epsilon^{ac} +h^{ad} \left[ \mu^{-1} \right]^{cb} - h^{ac} \left[ \mu^{-1} \right]^{db}\right.\nonumber\\
		 	& + h^{bc} \left[ \mu^{-1} \right]^{ad} - h^{bd} \left[ \mu^{-1} \right]^{ac} + (h^{bd} h^{ac} - h^{bc} h^{ad}) \left[ \mu^{-1} \right]^e{}_e \nonumber\\
		 	&+\left. \eps^{fabe}(V^d \zeta_e{}^c - V^c \zeta_e{}^d)V_f + \eps^{fcde}(V^b \zeta_e{}^a - V^a \zeta_e{}^b)V_f\right).\label{eq:ZBel3}
 	\end{align}
 	Note that every term involving two $V$'s corresponds to a time-projection.
 	\end{widetext}
 	
 	The three equations~\eqref{eq:ZBel1}, \eqref{eq:ZBel2} and \eqref{eq:ZBel3} now are the Bel decomposition of the constitutive tensor. They show that, once an observer's four-velocity $V^a$ is chosen, there exists a unique decomposition of $Z$ into three constitutive matrices $\epsilon, \mi$, and $\zeta$ for that given observer. Put differently, this decomposition clearly shows the observer-dependence of $\epsilon, \mi$, and $\zeta$.
 	
 	While $\epsilon$ and $\mi$ are automatically symmetric, $\zeta$ has (a priori) no symmetries. Also note that the antisymmetry properties of either $\eps^{abcd}$ or $Z^{abcd}$ guarantee that $\epsilon, \mi$, and $\zeta$ are four-orthogonal to $V$:
 	\begin{equation}
	 	\epsilon^{ab} V_b = \mi^{ab} V_b = \zeta^{ab} V_b = \zeta^{ba} V_b = 0.
 	\end{equation}
 	
 	It should be mentioned that, ironically, the names given to the three independent matrices $W_\epsilon$, $W_\zeta$, and $W_\mu$ encountered in this decomposition \emph{in the GR community} are very misleading in the present context: In GR, the Bel decomposition of the Riemann tensor is used to find dynamical analogies between the Einstein equations on the one hand, and the Maxwell equations on the other hand. In our case, now, the role of the Bel decomposition is only kinematical and entirely in the realm of electromagnetism itself. For example, what goes under the name of \enquote{electric tensors} in \cite{ChoBru} corresponds to both the permittivity and the permeability tensors, while the \enquote{magnetic tensors} here are the magneto-electric tensor and its transpose.
 	
 	\section{Moving isotropic media}\label{sec:iso}
 	An isotropic medium with no magneto-electric effects moving with four-velocity $V^a$ has \emph{in its rest frame} permittivity tensor and permeability tensor given by the following equations:
 	\begin{subequations}
 		\begin{equation}
	 		\epsilon^{ab} = \epsilon (g^{ab}+V^a V^b ) = \epsilon h^{ab} 
 		\end{equation}
 		and
 		\begin{equation}
	 		[\mu^{-1}]^{ab} = \mu^{-1} (g^{ab}+V^a V^b ) = \mu^{-1} h^{ab}.
 		\end{equation}
 	\end{subequations}
 	Inserting this in the Bel-decomposed constitutive tensor $Z^{abcd}$ yields, according to equation~\eqref{eq:ZBel2},
 	\begin{align}
 		Z^{abcd} = -{\epsilon\over2}& (  V^a V^c h^{bd} + V^b V^d h^{ac} -V^a V^d h^{bc}\nonumber\\& - V^b V^c h^{ad}) + {\mu^{-1}\over2} ( h^{ac} h^{bd} - h^{ad} h^{bc}).
 	\end{align}
 	This in turn can be rearranged to get
 	\begin{align}
	 	Z^{abcd} = {\frac{\mu^{-1}}{2}} &\kle{ \left(h^{ac} - \epsilon\mu V^a V^c\right) \left(h^{bd} - \epsilon \mu V^b V^d\right)\right.\nonumber\\
	 	&\left.- \left(h^{ad} - \epsilon\mu V^a V^d\right) \left( h^{bc} - \epsilon\mu  V^b V^c\right)},\label{eq:Ziso}
 	\end{align}
 	which then in turns lends itself to two different applications: The first is to derive again the consistency condition~\eqref{eq:conscond3+1}. The second is to get fully covariant expressions for the magneto-electric effect of moving media. We shall do both consecutively in the following short subsections.
 	
 	\subsection{The consistency condition}
 	Taking from equation~\eqref{eq:actioneff} that an effective metric would mean
 	\begin{equation}
 		Z^{abcd} = \sqrt{\frac{\det (\g)}{\det (g)}}\kl{\gi^{ac}\gi^{bd} - \gi^{ad}\gi^{bc}},
 	\end{equation}
 	and comparing this with the just derived equation~\eqref{eq:Ziso}, we see that the existence of an effective metric $\g$ would imply
 	\begin{equation}
 		\sqrt[4]{\det(\g)\over\det(g)} \;  \gi^{ab}  = \mu^{-1/2} \left(h^{ab} - \epsilon\mu V^a V^b\right).
 	\end{equation}
 	Taking determinants on both sides, we get the following equivalent of the previously derived consistency condition~\eqref{eq:conscond3+1} in the special case of an isotropic medium:
 	\begin{equation}
 		-1 = -\frac{\epsilon}{\mu}.
 	\end{equation}
 	If the isotropic medium fulfils this condition we can then immediately write down the inverse effective metric as
 	\begin{equation}
 		\gi^{ab}  \propto \left(h^{ab} - \epsilon\mu V^a V^b\right)
 	\end{equation}
 	or more specifically as
 	\begin{equation}
 		\gi^{ab}  =  (\epsilon\mu)^{-1/4} \left(h^{ab} - \epsilon\mu V^a V^b\right).
 	\end{equation}
 	
 	\subsection{The magneto-electric effect of moving media}
 	Instead of looking for the possibility for an effective metric describing the constitutive tensor, we can also use the results of section~\ref{sec:frames} to see what \enquote{constitutive matrices} an observer, who is not comoving to the natural reference frame of the medium, would measure. To this end, let us look at the equations~\eqref{eq:consmatW}, again, with $W^a$ denoting the four-velocity of the observer. First, we shall calculate the permittivity $\epsilon_W^{ab}$. After some algebra equation~\eqref{eq:epsW} is evaluated to be
 	\begin{align}
 		\epsilon_W^{bd} =& -2 Z^{abcd} W_a W_c,\\
 		=& \mu^{-1} (g^{bd}+W^b W^d) + (\epsilon-\mu^{-1}) \left( g^{bd} (V\cdot W)^2 \right.\nonumber\\&\left.- (W^b V^d+V^b W^d)(V\cdot W) - V^b V^d\right).
 	\end{align}
 	Defining
 	\begin{equation}
 		h_W^{bd} \defi g^{bd} + W^b W^d,
 	\end{equation}
 	and realizing that
 	\begin{align}
 		h_W^{be} h_{ef} h_W^{fd} =& g^{bd} + [1+(V\cdot W)^2] W^b W^d \nonumber\\&+ (V\cdot W) [W^b V^d + V^d W^b] + V^b V^d,
 	\end{align}
 	we can even simplify $\epsilon_W^{ab}$ further to 
 	\begin{subequations}
 		\begin{align}
	 		\epsilon_W^{bd} 
	 		=&  \mu^{-1} (h_W^{bd}) \nonumber\\
	 		&-(\epsilon-\mu^{-1})\left[ h_W^{be} h_{ef} h_W^{fd}\right. \left.-  [1+(V\cdot W)^2] h_W^{bd}\right],\\
	 		=&\epsilon \; h_W^{bd}  +(\epsilon-\mu^{-1})  \kle{(V\cdot W)^2 h_W^{bd}    - h_W^{be} h_{ef} h_W^{fd}}.\label{eq:epsiso}
 		\end{align}
 	\end{subequations}
 	
 	For $[\mu_W^{-1}]^{bd}$ it is helpful to realize that $h^{ab}-\epsilon\mu V^a V^b$ is for the following calculational needs the inverse of a (Lorentzian) metric $\mathcal{G}_{ab}$.\footnote{On a purely formal level it is of the form of the inverse Gordon metric\cite{lrrAnalogue}, even though at this stage we have not yet imposed the consistency condition which may or may not hold. And given most materials' properties it most likely will not! On the other hand, the Gordon metric \emph{does} have general validity in the ray optics limit, as opposed to wave optics.} Therefore, it will have an associated Levi-Civita tensor (density) $\eps^\mathcal{G}$. This then means that we can \enquote{pictorially} --- meaning we forget numerical factors and physical coefficients like $\mu^{-1}$ --- rewrite the defining equation~\eqref{eq:muW} to showcase the tensorial dependencies:
 	\begin{widetext}
 	\begin{subequations}
 		\begin{align}
 			\kle{\mu_W^{-1}}^{\bullet\bullet} &\simeq \kl{\ast \kl{\mathcal{G}^{-1}\mathcal{G}^{-1} - \mathcal{G}^{-1}\mathcal{G}^{-1}}\ast }^{\bullet\bullet\bullet\bullet}W_\bullet W_\bullet,\\ 
 			&\simeq \eps^{\bullet\bullet}{}_{\bullet\bullet} \kl{\mathcal{G}^{-1}\mathcal{G}^{-1} - \mathcal{G}^{-1}\mathcal{G}^{-1}}^{\bullet\bullet\bullet\bullet}\eps^{\bullet\bullet}{}_{\bullet\bullet}W_\bullet W_\bullet,\\
 			&\simeq [g^{-1}]^{\bullet\bullet}[g^{-1}]^{\bullet\bullet}[g^{-1}]^{\bullet\bullet}[g^{-1}]^{\bullet\bullet} \underbrace{\eps_{\bullet\bullet\bullet\bullet}\eps_{\bullet\bullet\bullet\bullet}}_{\simeq \sqrt{\frac{\det g}{\det \mathcal{G}}}^2\; \eps^\mathcal{G}_{\bullet\bullet\bullet\bullet}\eps^\mathcal{G}_{\bullet\bullet\bullet\bullet}}\kl{\mathcal{G}^{-1}\mathcal{G}^{-1} - \mathcal{G}^{-1}\mathcal{G}^{-1}}^{\bullet\bullet\bullet\bullet} W_\bullet W_\bullet,\\
 			&\simeq \frac{\det g}{\det \mathcal{G}} [g^{-1}]^{\bullet\bullet}[g^{-1}]^{\bullet\bullet}[g^{-1}]^{\bullet\bullet}[g^{-1}]^{\bullet\bullet} \underbrace{\eps^\mathcal{G}_{\bullet\bullet\bullet\bullet}\eps^\mathcal{G}_{\bullet\bullet\bullet\bullet}\kl{\mathcal{G}^{-1}\mathcal{G}^{-1} - \mathcal{G}^{-1}\mathcal{G}^{-1}}^{\bullet\bullet\bullet\bullet}}_{\simeq \kl{\mathcal{G}\mathcal{G} - \mathcal{G} \mathcal{G}}_{\bullet\bullet\bullet\bullet}} W_\bullet W_\bullet,\\
 			&\simeq \frac{\det g}{\det \mathcal{G}} \kl{\kle{g^{-1}\mathcal{G}g^{-1}}\kle{g^{-1}\mathcal{G}g^{-1}}-\kle{g^{-1}\mathcal{G}g^{-1}}\kle{g^{-1}\mathcal{G}g^{-1}}}^{\bullet\bullet\bullet\bullet}.
 		\end{align}
 	\end{subequations}
 	\end{widetext}
 	Now $\frac{\det g}{\det \mathcal{G}}$ evaluates to $\epsilon\mu$ and
 	\begin{equation}
 		\kle{g^{-1}\mathcal{G}g^{-1}}^{\bullet\bullet} = g^{\bullet\bullet} + \kl{1- \ed{\epsilon\mu}} V^\bullet V^\bullet.
 	\end{equation}
 	With this we can then perform a similar analysis to the one for $\epsilon^{ab}$ and arrive at
 	\begin{equation}
	 	[\mu^{-1}_W]^{bd}  = \frac{h_W^{bd}}{\mu}  +(\mu^{-1}-\epsilon)  \kl{ (V\cdot W)^2 h_W^{bd}    - h_W^{be} h_{ef} h_W^{fd}}.\label{eq:muiso}
 	\end{equation}
 	Finally, starting from equation~\eqref{eq:zetaW} we arrive, again after some algebra, at the equation
 	\begin{equation}
 		\zeta_W^{ac} = (\epsilon-\mu^{-1}) (V\cdot W) \kl{\epsilon^{acef}W_e V_f} \label{eq:zetaiso}
 	\end{equation}
 	for the magneto-electric matrix $\zeta_W^{ac}$.
 	
 	Note that this calculation reproduces several important physical insights:
 	\begin{enumerate}
 		\item If we pull out a factor $\epsilon$ in front of the right-hand side of equation~\eqref{eq:zetaiso}, the remainder of the right-hand sides will contain a factor of $1-\nicefrac{1}{\epsilon\mu} = 1-\ed{n^2}$ --- which nicely reproduces the Fresnel--Fizeau effect in flat space.
 		\item Similarly, in flat space and if both the observer and the natural reference frame of the medium are inertial frames, note that $(V\cdot W)^2 = \gamma^2$ is just the Lorentz factor we expect second-rank tensors like the \enquote{constitutive matrices} to have.
 		\item Finally, equation~\eqref{eq:zetaiso} gives the well-known result that a moving medium will have magneto-electric effects, even if it would not at rest. Again, this is tightly related to the Fresnel--Fizeau effect, but is a more general result.
 		\item Also, isotropy is lost under a change of observer. This happens even for inertial observers in Minkowski space and is intimately connected to the appearance of magneto-electric effects.
 	\end{enumerate}

 \end{document}